\documentclass[letterpaper, a4paper]{amsart}

\usepackage{graphicx}
\usepackage{latexsym}
\usepackage[all]{xy}
\usepackage{color}

\usepackage[margin=1.5in]{geometry}

\numberwithin{equation}{section}
\begin{document}
\begin{title}[Enhancement of the range-delay accuracy]
 {Enhancement in the mean square range-delay accuracy by means of multiple entangled photon states quantum illumination}
\end{title}
\date{\today}
\maketitle
\thispagestyle{empty}
\begin{center}
\author{Ricardo Gallego Torrom\'e\footnote{Email: rigato39@gmail.com}}
\end{center}
\begin{center}
\address{{\small Fraunhofer Institute for High Frequency Physics
and Radar Techniques FHR,\\
 Fraunhoferstraße 20 | 53343 Wachtberg | Germany.}\\
\address{Department of Mathematics,
University of Primorska, Koper, Slovenia}}
\end{center}
\bigskip
\begin{abstract}
It has been discussed recently how quantum illumination can be used to increase the accuracy of the value range-delay measurement \cite{Zhuang Shapiro 2022} in the domain of SNR compatible with current radar systems. However, the advantage described in \cite{Zhuang Shapiro 2022} requires of a large integration time.
In this work it is argued that multiple entangled photon quantum illumination could help to reduce the integration time when evaluating range-delay. Our analysis is performed in the framework of three entangled photon states discrete quantum illumination protocols. In this setting it is shown explicitly that using a direct generalization of Lloyd's protocol to the case where signal states describe two photons presents interesting advantages: 1. The reduction of the probability of error with respect to Lloyd's quantum illumination, 2. The reduction of the integration time with respect to Lloyd's quantum illumination, 3. The increase in the ultimate in the measurement of the range-delay accuracy.
\end{abstract}
\section{Introduction}
In a recent article, Q. Zhuang and J. H. Shapiro have illustrated how quantum illumination offers an enhancement of the range-delay measurement accuracy, which is manifestly large in a domain of signal-to-noise ratio (SNR) compatible with radar applications \cite{Zhuang Shapiro 2022}. Quantum illumination (QI) implies an enhancement in SNR with respect to thermal illumination \cite{Lloyd2008} or, when using Gaussian quantum illumination, with respect to coherent light illumination (CI), the benchmark of non-entangled states illumination \cite{Tan,Pirandola et al. 2018,Shapiro2019,GallegoBenBekhtiKnott2020}.

Zhuang and Shapiro have found how QI improves the range-delay measurement accuracy in two different forms. For a quantum pulse-compression radar of chord type with high time-bandwidth signal and high time-bandwidth idler, the advantage in the SNR offered by QI is translated in a reduction of the {\it threshold SNR} where the {\it Cramers-Rao bound} (CRB) for the range-delay is sub-optimal and instead, it is more convenient to consider the {\it Ziv-Zakai error bound} \cite{Zakai Ziv,Chow Schultheiss,Weiss Weinstein 1983,Weiss Weinstein 1984,J. Ianniello Weinstein Weiss, Tsang 2012}. It is when one considers the more optimal Ziv-Zakai bound that the advantage on the mean square range-delay accuracy when using QI is manifested in two ways. First, a generic reduction of the error in the determination of the range-delay, that happens even before the threshold SNR. Second, the analysis of Zhuang and Shapiro also showed a large advantage in the mean square range-delay accuracy when using quantum illumination with respect to the use of coherent states illumination that can be up to several dozen dB at the SNR threshold. This was attributed to the intrinsic non-linearity of the problem of the range-delay performance with at different SNR.

There is an intrinsic difficulty to exploit the aforementioned advantage, since the pulse duration required must be large compared with the scales of the integration time in standard radar systems. For practical situations, comparable to standard radar application conditions, the benefit in range-delay measurement accuracy is obtained for integration times of the order of $10^{2}\,s$. This is because the SNR required to exploit the non-linear effect discussed is not small, typically for microwave radar applications, of the order of $5$ to $10$ dB. For the type of pulse-compression signal discussed in \cite{Zhuang Shapiro 2022}, an estimate of the required SNR is given by the expression
$SNR=\,\frac{\kappa\,M\,N_S}{N_B},$ where $\kappa$ stands for a single round trip radar-to-target-radar transmissivity and it is assumed constant, $M$ is the time-bandwidth product, $N_S$ is the average number of photons per mode and $N_B$ is the background average number of photons per mode. In the microwave regime, $N_B$ is large (of the order $10^2$ or more at room temperatures) and $N_S$ is small (of order $10^{-2}$). Typically $\kappa$ is small too compared with $1$. Moreover, quantum enhancement of SNR (and hence the benefits) happens for $M$ very large. However, there are limitations in the time-bandwidth product $M$ due to the mechanism of the generation of the entangled states, which is usually spontaneous parametric down conversion (SPDC). Hence a natural mechanism to increase $M$ (the time-bandwidth product) is to increase the time pulse $T$. This is one of the causes of the problem of large integration time scale mentioned above.

\begin{figure}
\graphicspath{{C:\Users\rgall\Desktop\Ricardo 09-2019\Desktop\Ricardo backup 2018 05\Ricardo\trabajos mios, PhDs y tesinas/}}
\begin{center}
\includegraphics[scale=0.3]{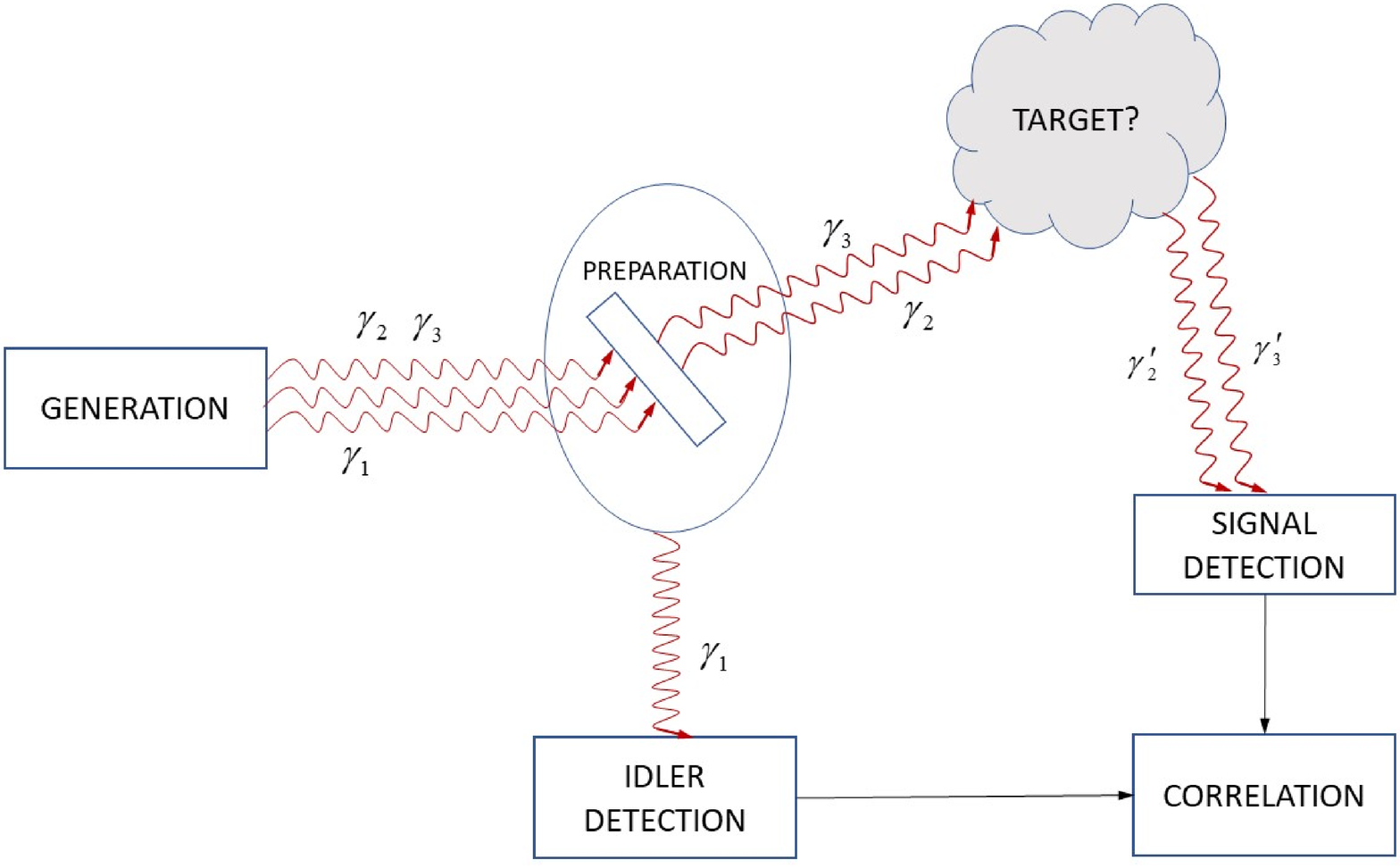}
\caption{{\small {\bf Quantum illumination with three entangled photon states.} }}
\label{qi2p direct measurement}
\end{center}
\end{figure}

In this work it is explored the possibility to use quantum illumination with states describing multiple entangled photons to increase the accuracy in the range-delay determination by radar protocol and to reduce the required integration time. The protocol of  quantum  illumination with three entangled photon states is illustrated in figure \ref{qi2p direct measurement} and can be described in a nutshell as follows \cite{Ricardo 2021}. Three photons correlated in time and in frequency are generated as a result of a four photon interaction in a non-linear system \cite{GarrisonChiao}. Although the states differ from the type described in this paper, experimental demonstrations of three photon states generation in the microwave can be found for instance in \cite{C.W. Sandbo Chang et al. 2020}. Such demonstration shows that quantum illumination with multiple entangled states is potentially feasible experimentally. Thus assuming three entangled photon states, for each of the entangled three photon systems, two photons (signal photons) are sent to explore a region where the target could be located. One photon is either retained or directly detected just after generation (idler photon), depending on the specific protocol used. When two photons are detected simultaneously using direct detection methods, then the temporal correlation between the idler photon and the two signal photons is checked and if the correlation test results positive, it is claimed that there is a detection event.
Then the experiment is repeated $m>>1$ independent times, in order to discriminate between the case $H_0$ corresponding to the situation where the target is not there and the case $H_1$ that corresponds to the situation where the target is there.
It will be shown that Lloyd's quantum illumination with multiple entangled photons implies an increase in the accuracy of the range time delay determination and a reduction of the integration time for the protocol of range-delay measurement when comparing with Lloyd's quantum illumination.

The analysis and estimates described below compare the low noise model of Lloyd discussed first in \cite{Lloyd2008} and the model of three entangled photon states quantum illumination discussed in \cite{Ricardo 2021}, which is a direct generalization of LLoyd's model. The comparison of these two models shows a reduction of the corresponding probability of errors. Such reduction is translated, for the models of quantum illumination discussed in the test, in an enhancement of the accuracy in the delay time delay measurement error in terms of the associated Ziv-Zakai error bounds in the regime of low SNR. A second benefit is a reduction of the integration time when using multiple entangled photon states illumination. These benefits are found for any values of the SNR, as long as it is kept small.

We remark here that the domain of definition of the models discussed in this paper is different than the one discussed in \cite{Zhuang Shapiro 2022}. Here we have considered low noise environment, while in the case considered by Zhuang and Shapiro is the situation of a noise environment $N_B>>1$. Also, the type of models considered are different: while we have used discrete models for the quantum illumination as envisaged Lloyd in \cite{Lloyd2008} and its direct generalization in \cite{Ricardo 2021}, in \cite{Zhuang Shapiro 2022} the models consistend of quantum pulse-compression radar subjected of Gaussian entanglement states, where Gaussian quantum illumination enhancement is applicable. Realizing these differences illustrates further the objective of the present work, which was to show the potential advantage of multiple entangled states for the enhancement in the accuracy of the range-delay measurement.

\section{Lloyd's quantum illumination with multiple entangled photons signals beams} \label{Muliple entangled photons}
We consider in this paper the situation where the back-ground noise is low, with average number of photons per mode $N_B <<1$ is small and the time-bandwidth product $M$ is large compared to $1$, as in the original work of Lloyd \cite{Lloyd2008}.
We specialize further our considerations to the case initial state of three entangled photons is of the form
\begin{align}
|\psi\rangle_3 =\,\frac{1}{\sqrt{M}}\,\sum^{M}_{\alpha=1}\,\hat{a}^\dag(\omega_1(\alpha),\vec{k}_1(\alpha))\,\hat{a}^\dag(\omega_2(\alpha),\vec{k}_2(\alpha))\,\hat{a}^\dag(\omega_3(\alpha),\vec{k}_3(\alpha))\,|0\rangle,
\label{three generation}
\end{align}
where $\alpha =1,...,M$ indicates the number of modes of the states and coincides with the time-bandwidth product.
The photons $1$, $2$, $3$ are correlated in time in the sense that they are simultaneously generated, in frequency, by the relation $\omega_0=\,\omega_1+\omega_2+\omega_3$ and in momentum, by the relation
$\hbar\vec{k}_0=\,\hbar\vec{k_1}\,+\hbar\vec{k}_2\,+\hbar\vec{k}_3 $  (see for instance, \cite{GarrisonChiao}, section 13.4), where $(\omega_o,\hbar\vec{k}_0)$ are the frequency and momentum of the pump beam.
The initial state can be transformed by an unitary operator $\mathcal{U}=\,I_1\otimes I_2\otimes U_3$ that sends $\frac{\vec{k}_3}{\|\vec{k}_3\|}$ to $ \vec{k}'=\frac{\vec{k}_2}{\|\vec{k}_2\|})$ to a state of the form
\begin{align}
\widetilde{|\psi\rangle}_3 =\,\frac{1}{\sqrt{M}}\,\sum^{M}_{\alpha=1}\,\hat{a}^\dag(\omega_1(\alpha),\vec{k}_1(\alpha))\,\hat{a}^\dag(\omega_2(\alpha),\vec{k}_2(\alpha))\,\hat{a}^\dag(\omega_3(\alpha),\vec{k}'_2(\alpha))\,|0\rangle,
\label{three generation collimated}
\end{align}
where two of the photons have parallel momenta $\vec{k}_2 \, \| \,\vec{k}'_2$, but were the frequencies $\omega_2$ and $\omega_3$ can be different. After this preparation,
the beam is split into two beams: the idler beam, which is composed by photons with momentum $(\omega_1,\vec{k}_1)$, and the signal beam, which is composed by states with two photons with momentum $(\omega_3,\,\|\vec{k}_3\|\,\vec{k}'_2)$.
The signal and the idler state are not any more entangled, but mixed states. However, the following correlations persists:
\begin{itemize}
\item The time correlation in the generation of the three photons,

\item The photons on each state exhibit correlations in energy,
\begin{align}
\hbar\omega_0=\,\hbar\omega_1(\alpha)+\,\hbar\omega_2(\alpha)+\,\hbar\omega_3(\alpha),\quad \alpha =1,...,M
\label{correlation relation 1}
\end{align}
\end{itemize}
 Note that although the photons are initially correlated in momenta, such correlation is lost during the preparation of the signal and idler beams. However, The correlations in energy and time remain after the preparation.

Adopting the above framework, one declares that the target is present if two photons in the spectrum range of the signal, are detected simultaneously with an idler photon in a joint measurement of the idler and signal beam and if the energy constraints \eqref{correlation relation 1} holds good. This criterion is formally implemented by adopting a dichotomic positive operator valued measure (POVM) $\{\tilde{\Pi}_0,\tilde{\Pi}_1\}$, where the operator $\Pi_0$ corresponds to the negative detection of two simultaneous photons and the operator $\Pi_1=\,\mathbb{I}-\Pi_0$ corresponds to the positive detection of two simultaneous photons such that the constraints \eqref{correlation relation 1} hold good.

\subsection{Enhancement of the signal to noise ratio with respect to classical light illumination}
In the following paragraphs the probability of error and the SNR for non-entangled illumination and for quantum illumination with multiple entangled photon signal states and one photon idler in the case $N=2$ is evaluated. The procedure that we follow is analogous to the original treatment of S. Lloyd \cite{Lloyd2008}. Although we are here interested in quantum illumination protocols, it is instructive to follow a complete treatment similar to the original Lloyd's theory, since it will help to motivate the diverse density matrices that we will introduce. Therefore, we distinguish two protocols:
\\
{\bf A. Illumination with non-entangled light}. When no target is present and the illumination is made by means of non-entangled light, the quantum state is described by the product density matrix
\begin{align*}
\tilde{\rho}_0=\,\rho_0\otimes \,\rho_0,
\end{align*}
where $\rho_0$ is the density matrix for a thermal background state with average number per mode $N_B$. In the case where $N_B<<1$, the state $\rho_0$ is given approximately by the expression \cite{Marco Lanzagorta 2011}
\begin{align}
\rho_0  \approx \left\{(1-\,M\,N_B)|0\rangle\langle 0|+\,N_B\,\sum^M_{k=1}\,|k\rangle \langle k \right\},
\label{state for noise}
\end{align}
where $\{|k\rangle\}$ is a number photon state basis.
 The probability of false positive is read directly from the structure of the state and, due to the criteria of detection discussed above, it is associated with the event of detecting two signal photons in the same time window. In the case of a low bright environment ($N_B\ll\,1$), the probability of false positive is determined by the criteria of detection of two independent photons simultaneously with the required energies $\omega_2$, $\omega_3\,\in Spect$ and is approximated in the regime $N_B\ll \,1$ by the expression
\begin{align}
\tilde{p}_0(+)\approx\,(N_B)^2,
\label{N=3 non-entangled light nothere}
\end{align}
where it is assumed that the detections of each of the two photons are statistically independent events.

If the target is present, then the state is given by the density matrix
\begin{align*}
\tilde{\rho}_1& =\,(1-\eta)\,\rho_0\otimes\,\rho_0+\eta\tilde{\rho}\\
& \approx (1-\eta)\left\{(1-\,M_B\,N_B)|0\rangle\langle 0|+N_B\sum^{M_B}_{\beta=1}\, a^\dag ({k}(\beta))|0\rangle\,\langle 0|a({k}(\beta))\right\}^2+\eta\tilde{\rho} ,
\end{align*}
where $\tilde{\rho}$ stands for the density matrix of state describing two signal photons when using classical illumination and $\eta$ is the reflectivity of the target.
The probability for detecting the arrival of two independent photons simultaneously takes the form
\begin{align}
\tilde{p}_1(+)=\,(1-\eta)N^2_B \,+\eta.
\end{align}
The probability of false negative is the of the form
\begin{align}
\tilde{p}_1(-)=\,1-\left((1-\eta)N^2_B \,+\eta\right).
\end{align}

Analogously as in the case of illumination with one photon state \cite{Lloyd2008,Marco Lanzagorta 2011}, the SNR in quantum illumination with two signal state photons when the illumination is performed with non-entangled light is given by the expression
\begin{align}
SNR_{CI2P}=\,\frac{\tilde{p}_1(+)}{\tilde{p}_0(+)}\approx \,\frac{(1-\eta)N^2_B+\eta}{N^2_B}.
\label{SNR noentanglement in Maccone-Ren illumination}
\end{align}
\\
{\bf B. Quantum illumination with three photon entangled states}.
When there is no target present, the  noise-idler system is described by a density matrix $\tilde{\rho}^e_0$ of the form
\begin{align}
\tilde{\rho}^e_0 \approx \,\left\{(1-\,M_B\,N_B)|0\rangle\langle 0|+\,N_B\,\sum^{M_B}_{\beta=1}\,a^\dag (\omega(\beta),\vec{k}(\beta))|0\rangle\,\langle 0|a(\omega(\beta),\vec{k}(\beta))\right\}^2\otimes\,\tilde{\rho}_I,
\end{align}
where $\tilde{\rho}_I$ is the idler state given by the expression
\begin{align}
\tilde{\rho}_I=\,\frac{1}{M}\,\sum^M_{\alpha =1}\,|{k}_1(\alpha)\rangle\langle {k}_1(\alpha)| .
\label{idler state}
\end{align}

The probability of a false positive is the probability attributed to the presence of the target by the detection of two simultaneously returned noise photons. Within the scope of the approximations that we are considering, the probability of a false positive is independent of the details of the signal state and given by the expression
\begin{align}
\tilde{p}^e_0(+)\approx\,\left(\frac{N_B}{M}\right)^2.
\label{probability false positive 3p}
\end{align}
This is just the square of the probability of false positive in Lloyd's theory (equation \eqref{entanglemet0+}). It is the probability to detect two photons from the background.

When the target is present, the state after considering decoherence and signal-target interaction is of the form
\begin{align}
\tilde{\rho}^e_1 =\,(1-\eta)\cdot \tilde{\rho}^e_0+\,\eta\,\tilde{\rho}^e,
\label{density2qi}
\end{align}
where $\tilde{\rho}^e$ is the density matrix describing the joint system signal photons/idler after mixing with the noise environment and scattering with the target has happened, processes that induce the transformation from the original state $ |\Phi\rangle$  to the state $\tilde{\rho}^e$. 
Adopting the state \eqref{density2qi}, the probability of positive detection when using entangled states is of the form
\begin{align}
\tilde{p}^e_1(+)=\,(1-\eta)\,\left(\frac{N_B}{M}\right)^2+\eta .
\end{align}
The probability of a false negative is
\begin{align}
\tilde{p}^e_1(-)=\,1-\left((1-\eta)\,\left(\frac{N_B}{M}\right)^2+\eta\right).
\label{probability of false negative 3p}
\end{align}
The signal to noise ratio for quantum illumination with signal states described by two photons is of the form
\begin{align}
SNR^e_{QI2P}=\,\frac{\tilde{p}^e_1(+)}{\tilde{p}^e_0(+)}\approx\,\left(\frac{M}{N_B}\,\right)^2\,\left((1-\eta)\left(\frac{N_B}{M}\right)^2+\eta\right) .
\label{multiple photon entangle illumination}
\end{align}
This expression compared with the SNR for Lloyd's model shows an enhancement in the SNR when considering states with $N=2$ photons signal states entangled with a $1$-photon state idler.
\subsection{Probability of error for Lloyd's type protocols}
We show now that by using $N=2$ instead than $N=1$ photon signal states for Lloyd's type quantum illumination produce a reduction in the probability of error.
In the problem of discrimination between two quantum states $\tilde{\rho}_0$ and $\tilde{\rho}_1$ where one has associated an a priori probability $p_0$ and the other $p_1$ such that $p_0+p_1=1$, the probability of error is given by the expression
\begin{align}
P^{err}=\,p_0\,p(H_1|\tilde{\rho}_0)+\,p_1\,p(H_0|\tilde{\rho}_1),
\label{definition of probability of error}
\end{align}
where $p(H_i|\tilde{\rho}_j)$ is the conditional probability that one decides by the hypothesis $H_i$ when in fact the state is $\tilde{\rho}_j$. The minimum probability of error in the discrimination of the states $\tilde{\rho}_0$ and $\tilde{\rho}_1$ is achieved by the Helstrom's bound for certain selection of the POVM operators \cite{Helstrom 1976}. In the case of Lloyd's quantum illumination protocol, it was proved that the measurement scheme proposed in \cite{Lloyd2008} achieves the Chernoff's bound, that is asymptotically tied to the Helstrom's bound. Note that when considering the particular measurement in terms of two photon detection scheme discussed above by adopting the operators $\tilde{\Pi}_0$, $\tilde{\Pi}_1$, we are theoretically constrained to a particular subset of POVM.  However, our objective is to compare Lloyd's type quantum illumination protocols with $N=1$ and with $N=2$ signal photon states. For this goal it will be enough to show enhancement of the case $N=2$ with respect to the protocol $N=1$ if we can show enhancement when evaluation the probability of error for the case $N=2$ when using POVM $\{\tilde{\Pi}_0$, $\tilde{\Pi}_1\}$ with respect to the probability of error standard Lloyd's theory with $N=1$.

Let us consider the symmetric scenario $p_0=\,p_1 =1/2$. The probability of error takes the form
\begin{align*}
P^{err}=\,\frac{1}{2}\left(\tilde{p}^e_0(+)+\tilde{p}^e_1(-) \right).
\end{align*}

 In the regime of low environmental noise $N_B<<1$ and high time-bandwidth product $M>> 1$. For Lloyd's quantum illumination protocol as discussed in \ref{AppendixLloyd quantum illumination}, the probability of error is given by the expression
\begin{align}
{p}^{err}_{1PQI}=\,\frac{1}{2}\left(1+\eta\,\left(\frac{N_B}{M}\right)-\eta\right).
\label{probability of error for N=1}
\end{align}
The probability of error in the case of entangle states with $N=2$ can be evaluated from the expressions \eqref{probability false positive 3p} and \eqref{probability of false negative 3p},
\begin{align}
{p}^{err}_{2PQI}\leq \,\frac{1}{2}\left(1+\eta\,\left(\frac{N_B}{M}\right)^2-\eta\right).
\label{probability of error for N=2}
\end{align}
The inequality character of the relation \eqref{probability of error for N=2} is justified because the expression obtained is associated to a particular class of POVM $\{\tilde{\Pi}_0$, $\tilde{\Pi}_1\}$ as discussed above. To find the exact expression for ${p}^{err}_{2PQI}$ either the Helstrom's bound or the corresponding Chernov's bound must be evaluated. However, for the purposes of showing enhancement of the case $N=2$ with respect to the case $N=1$, the inequality relation is enough. Indeed, according to the expressions \eqref{probability of error for N=1} and \eqref{probability of error for N=2} it is clear that $\tilde{p}^e_{2PQI}<\,\tilde{p}^e_{1PQI}$. This is because
\begin{align*}
\left(1+\eta\,\left(\frac{N_B}{M}\right)^2-\eta\right)<\left(1+\eta\,\left(\frac{N_B}{M}\right)-\eta\right).
\end{align*}
The small difference between the term proportional to $N_B/M$ and $(N_B/M)^2$ implies, when amplified by making $m$ different experiments, a substantial decrease in the probability of error in Lloyd's type quantum illumination with multiple entangled photons with respect to standard Lloyd's quantum illumination protocol.

\section{Enhancement in the mean square range-delay accuracy by means of multiple entangled photon states quantum illumination}
The finding that motivated this paper is the founding by Zhuang and Shapiro of the enhancement in the mean square range-delay accuracy due to the use of entangled states in the determination of the range. To settle down the measurement in the accuracy in the range-delay, we would proceed in an analogous way as in \cite{Zhuang Shapiro 2022}. The radar signal is described by a state of the form \eqref{three generation collimated} during a time pulse duration $T$ and bandwidth $\Delta \omega$ such that $T\,\Delta \omega =M$. The range of the target and the delay time of fly are related by $\tau =\, 2 R/c$, where $c$ is the speed of light. The range $R$ is unknown, but it is assumed  that $R\in\,[R_{\textrm{min}}, R_{\textrm{max}}]$ and restricted by the condition $\Delta R :=R_\textrm{max}-\,R_\textrm{min}\ll\, (R_\textrm{min}+\, R_\textrm{max})/2$. This setting implies that the delay time of fly $\tau \in\, [\tau_{\textrm{min}}, \tau_{\textrm{max}}]$ and that $\Delta \tau := \tau_\textrm{max}-\,\tau_\textrm{min}\ll (\tau_\textrm{min}+\, \tau_\textrm{max})/2$.

For low enough SNR, one can settle the case under the SNR threshold from the linear regime to the non-linear regime in the delay time determination problem. Then the quantum Ziv-Zakai error bound provides an estimate for the rms of $\tau$ that is better than the Cramers-Rao bound. For the case under consideration of the delay of range accuracy determination, the quantum Ziv-Zakai error bound for $\tau$ is of the form \cite{Tsang 2012,Zhuang Shapiro 2022},
\begin{align}
\left(\delta \tau_{ZZB}(N)\right)^2 = \, \int^{\Delta \tau}_0\,d\tau ' \tau ' \left(1-\frac{\tau'}{\Delta \tau}\right)\,\tilde{p}^e_{NPQI}(\tau '),
\label{Ziv-Zakai error expression}
\end{align}
where $\tilde{p}^e_{NPQI}(\tau ')$ is the probability of error when using Lloyd's quantum illumination with states with $N$ entangled signal photon states entangled with one photon idler state.

The first thing to notice is that the probability of error is constant in $\tau$, except at the initio and at the end of each pulse. Thus if we assume that $\tilde{p}^e_{NPQI}(\tau ')$ is constant in the integrand, then we have for $N=1$ and $N=2$ that the comparison between the corresponding probability of error leads to the expression and condition for a $m=1$ single detection experiment:
\begin{align}
\frac{\left(\delta \tau_{ZZB}(N=2)\right)^2}{\left(\delta \tau_{Ziv-Zakai error}(N=1)\right)^2} = \, \frac{\tilde{p}^e_{2PQI}}{\tilde{p}^e_{1PQI}}.
\label{enhancement in Ziv-Zakai error}
\end{align}
This relation evidences the existence of enhancement in the mean square range-delay accuracy when considering entangled light with $N=2$ photon signal states with respect to the case of $N=1$ photon signal states. We would like to remark that this is enhancement is a general feature, independent of the type of models used for quantum illumination, as long as ${\tilde{p}^e_{2PQI}}<{\tilde{p}^e_{1PQI}}$, at least under the domain below the SNB threshold. For the case of the models that we are considering in this paper, the expression for the probability of errors \eqref{probability of error for N=1} and \eqref{probability of error for N=2} leads to the relation
\begin{align*}
\frac{\left(\delta \tau_{ZZB}(N=2)\right)^2}{\left(\delta \tau_{ZZB}(N=1)\right)^2} =\, \frac{1+\eta\,\left(\frac{N_B}{M}\right)^2-\eta}{1+\eta\,\frac{N_B}{M}-\eta}<1,
\end{align*}
that implies a reduction in the Ziv-Zakai bound in the case of $N=2$.
\subsection{Integration time in quantum illumination with multiple entangled photons}
One relevant challenge for the practical use of the advantage found by Zhuang and Shapiro is the large time pulse required to translate the advantage in SNR in the enhancement of the range-delay accuracy at the SNR threshold. We aill argue that quantum illumination with multiple entangled states can help to reduce the integration time, at any value of SNR below the SNR threshold. 
For Lloyd's type quantum illumination protocols, multiple photon quantum illumination provides an advantage in SNR with respect to standard quantum illumination in the following sense \cite{Ricardo 2021}.
The SNR for $QI2R$ and for $QI$ are given respectively by the relations \cite{Ricardo 2021}:
\begin{align*}
& SNR^e_{QI2R}  =\,\left(\frac{M}{N_B}\,\right)^2\,\left((1-\kappa)\left(\frac{N_B}{M}\right)^2+\kappa\right),\\
& SNR^e_{QI} =\,\frac{M'}{N_B}\,\left((1-\kappa)\frac{N_B}{M'}+\kappa\right),
\end{align*}
where $M$ and $M'$ are a priori different time-bandwidth products. In the regime when $\kappa\ll\,1$, $M\gg\,1$, $M'\gg\,1$ and $N_B\ll 1$, the condition  $SNR^e_{QI2R}\approx\, SNR^e_{QI}$ requires that
\begin{align}
M\approx\,\sqrt{N_B\,M'}.
\label{condition for comparison M M'}
\end{align}
The relation \eqref{condition for comparison M M'} shows a reduction of the required time-bandwidth product for QIMP with respect to QI, here associated to the number of modes $M$ and $M'$ respectively. Thus if the bandwidths are the same, $\Delta \omega (1)=\,\Delta \omega (2)$ and assuming that $M=\,t(2)\,\Delta \omega (2)$ and $M'= \, t(1)\, \Delta \omega(1),$ then the corresponding integration times are related by the expression
\begin{align}
t(2)=\,\sqrt{N_B\,\frac{\,t(1)}{\Delta \omega(1)}},
\label{relation between the integration times}
\end{align}
where $t(1)$ is the integration time for quantum illumination with states describing a pair of one photon signal and one photon idler and $t(1)$ is the integration time for quantum illumination for states describing $2$ photons signal and one idler.

The relation \eqref{relation between the integration times} has been derived in the framework of Lloyd's quantum illumination models, where the condition of low noise $N_B<< 1$ is assumed. In this case. the relative reduction of the integration time is of the form
\begin{align}
\frac{t(2)}{t(1)}=\,\sqrt{\frac{N_B}{t(1)\Delta \omega }}
\label{reduction of the integration time}
 \end{align}
 with respect to the integration time of using standard Lloyd's quantum illumination. Thus for a bandwidth of $\Delta \omega =\,10^6 \,Hz$ and an integration time as large as $t(1)\sim \, 10^2\, s$ and not taking into account the factor $\sqrt{N_B}$, the reduction time factor is of order $10^{-2}$ when using three photon signal quantum illumination with respect to the usual LLoyd's quantum illumination protocol.

 \section{Discussion}
In this work we have argued that discrete quantum illumination with three, but in general with multiple $N>1$, entangled photon states describing idler-signal systems can reduce the integration time to increase the range-delay measurement. In particular, we have shown the existence of such enhancement in accuracy when using three photon states for Lloyd's type quantum illumination, namely, in a protocol of discrete quantum illumination where the signal state describes two photons system and the idler state describes one photon system, in the regime where the average number of noise photons per mode $N_B$ is very small compared with $1$ and the signal is small compared with the noise.

Since for the study of the accuracy of range-delay measurement at low SNR the Ziv-Zakai error provides a better bound than the Cramers-Rao bound \cite{Zhuang Shapiro 2022}, we have also compared the Ziv-Zakai error bound for the models of quantum illumination considered in this paper. Thus as an intermediate result, we have shown how under the conditions discussed here ($M$ large, $N_B<<1$, $N_S<<1)$ and for the models discussed above, the use of multiple entanglement states quantum illumination reduces the probability of error and increases the accuracy of the range-delay measurement. As a consequence, there is an effective reduction of the required integration time to perform the measurement. Note that the benefits discussed do not depend on the frequency range, being valid at the microwave, optical or X-ray regimes.

The regime where the models discussed in this paper apply differs from the regime discussed by Zhuang and Shapiro in \cite{Zhuang Shapiro 2022}, which was a bright environmental regime where $N_B\gg 1$. In order to confirm the existence of enhancements in practical situations in the microwave domain with $N_B>>1$, instead than using Lloyd's quantum illumination type models, a generalization of the theory developed by in Tan et al. for multiple photon states applicable for high bright environmental conditions  should be developed and the particular advantages discussed in this work extended to the noisy background case.

One of the findings discussed by Zhuang and Shapiro is that, due to the non-linear character of the problem range-delay determination, the advantage in SNR of Gaussian quantum illumination with respect to coherent light illumination can be translated to an enhancement in accuracy in the range-delay accuracy determination of the order of $10 \, dB$ or more at the SNR where the transition from linear to non-linearity happens. For the models of quantum illumination discussed in this work there is also enhancement in SNR \cite{Ricardo 2021}, we have evaluated the corresponding enhancements in SNR, probability of error an range-delay measurement accuracy, showing enhancement in sensitivity when multiple quantum illumination with respect to the usual Lloyd's quantum illumination. This advantage is also translated into an enhancement in accuracy by a reduction of the integration time. Differently than for the case investigated by Zhuang and Shapiro, our investigations are reduced to compare two different discrete quantum illumination protocols at low SNR and we were not able, by the methods used in the present work, to observe the decrease in the SNR threshold. However, this limitation is in any case, of relative relevance for our purposes, that was to demonstrate the potential advantages of multiple entangled states quantum illumination.

In the same line of constructive criticism it is worth good to remind that coherent illumination performs better than Lloyd's discrete quantum illumination \cite{ShapiroLloyd}. Therefore, although our study provides evidence of enhancement in accuracy in the range-delay determination when using multiple entangled photon states for quantum illumination, it does not provide an absolute comparison with respect to the classical benchmark.

  \appendix
  \section{Enhancement of sensitivity in Lloyd's quantum illumination: an illustrative example}\label{AppendixLloyd quantum illumination} In the following lines we discus in detail Lloyd's theoretical protocol \cite{Lloyd2008}. We partially follow the exposition described in \cite{Marco Lanzagorta 2011}.
\\
{\bf A. Non-entangled light illumination}. When the light used for experiments is described by non-entangled photons, the density matrix of the system idler-signal-noise, when the target is not there (hypothesis $0$) is
is given by the expression \eqref{state for noise}. Hence the probability of a false positive is
\begin{align}
p_0(+)=\,N_B,
\label{noentanglement0+}
\end{align}
 while the probability to be correct in the forecast that the target is not there is
\begin{align}
p_0(-)=1-\,p_0(+)=\,1-N_B.
\label{noentaglement0-}
\end{align}
If we repeat the experiment $m$ times, the probability of a false positive is
\begin{align*}
p_0(+,M)=\,(N_B)^m.
\end{align*}

If the target is there (hypothesis $1$), then the density matrix is given by
\begin{align}
\rho_1& =\,(1-\eta)\rho_0+\eta\tilde{\rho}\\
& \approx (1-\eta)\left\{(1-\,M N_B)|0\rangle\langle 0|+\,N_B\,\sum^M_{k=1}\,|k\rangle_n\langle k |_n\right\}+\,\eta \,|\psi\rangle_s\langle \psi|_s ,
\label{noentanglement 1 state}
\end{align}
where $|\psi\rangle_s$ stands for the state describing the signal, that one can assume first is a pure state and $\eta$ is the reflective index. It follows that the probability to measure the arrival of photon is
\begin{align}
p_1(+)=\,(1-\eta)N_B+\,\eta
\label{noentanglement1+}
\end{align}
 and that consequently, the probability of false negative is
 \begin{align}
 p_1(-)=\,1-p_1(+)=1-((1-\eta)N_B+\,\eta)=(1-\eta)(1-N_B).
 \label{noentanglement1-}
 \end{align}
 The signal to noise ratio is given by the expression
 \begin{align}
SNR_{QI}=\,\frac{p_1(+)}{p_0(+)}=\,\frac{((1-\eta)N_B+\,\eta)}{N_B}.
\label{SNR noentanglement}
\end{align}
\\
{\bf B. Entangled light illumination}. Let us now consider that the illumination is made using entangled states. For the case when there is no target there, the density matrix is given by the expression
\begin{align}
\rho^e_0 \approx \left\{(1-\,MN_B)|0\rangle\langle 0|+\,N_B\,\sum^M_{k=1}\,|k\rangle_n\langle k |_n\right\}\otimes\left(\frac{1}{M}\,\sum^M_{k=1}\,|k\rangle_A\langle k |_A \right),
\end{align}
where $\frac{1}{M}\,\sum^M_{k=1}\,|k\rangle_A\langle k |_A $ is the state of the idler.
The state
\begin{align*}
\rho_0=\,\left\{(1-\,MN_B)|0\rangle\langle 0|+\,N_B\,\sum^M_{k=1}\,|k\rangle_n\langle k |_n\right\}
\end{align*}
is the state that will describe the absence of the target. It determines the probability distributions to detect one photon due to noise only.
The modes determining the idler $k=1,...,M$ are selected to coincide with the modes of the noise. In this context, it is remarkable that the false positive probability for one individual detection,
\begin{align}
p^e_0(+)=\frac{N_B}{M}
\label{entanglemet0+}
\end{align}
 is dramatically reduced with the number of modes $M$. This was first highlighted by S. Lloyd in his seminal work \cite{Lloyd2008}. The probability of forecasting correctly the absence of the target  is given by the probability of the complement set,
 \begin{align}
 p^e_0(-)=\,1-\frac{N_B}{M}.
 \label{entanglemet0-}
 \end{align}
Note than when the experiment is repeated a number $m$ of times in a independent way, the probability of a false positive  after detecting $m$ independent photons is
\begin{align*}
p^e_0(+,m)=\,\left(\frac{N_B}{M}\right)^m.
\end{align*}

In the case that the target is there, for entangled states, the system idler-noise-signal is described by a density matrix of the form
\begin{align}
\rho^e_1 =\,(1-\eta)\cdot \rho^e_0+\,\eta\,\rho_s,
\end{align}
where $\rho_s$ is the density matrix of the signal photon system.
From this expression, one can extract the probability of detecting the target is
\begin{align}
p^e_1(+)=\,(1-\eta)\,\frac{N_B}{M}+\,\eta.
\label{entanglement1+}
\end{align}
The probability of no detection (interpreted as
a false negative) is of the form
\begin{align}
p^e_1 (-)=\,1-p^e_1(+)=\,(1-\frac{N_B}{M})\,(1-\eta).
\label{entanglement1-}
\end{align}
When applied $m$ independent experiments, the probability of right detection is
 \begin{align*}
p^e_1(+,m)=\,\left((1-\eta)\,\frac{N_B}{M}+\,\eta\right)^m
\end{align*}
For the case of false negative,
\begin{align*}
p^e_1 (-,m)=\,1-p^e_1(+)=\,(1-\frac{N_B}{M})^m\,(1-\eta)^m.
\end{align*}
\begin{align}
SNR^e_{QI}=\,\frac{p^e_1(+)}{p^e_0(+)}=\,\left(\frac{M}{N_B}\right)\,\left((1-\eta)\frac{N_B}{M}+\eta\right).
\label{SNR entanglement}
\end{align}
\\

\subsection*{Acknowledgements}
The author would like to thank the discussions with A. Froelich and M. Ummenhofer for relevant discussions on the topic of this paper. This work has been financially supported by the Fraunhofer Institute for High Frequency Physics and Radar Techniques.

\end{document}